\documentclass[a4paper,10pt]{article}
\usepackage[utf8x]{inputenc}
\usepackage{graphicx}
\usepackage{epsfig}
\usepackage[T1]{fontenc}
\usepackage{textcomp, amssymb}
\usepackage[utf8x]{inputenc}
\usepackage{amsmath}
\usepackage{amsfonts}
\usepackage{dsfont}
\usepackage{slashbox}
\usepackage{booktabs}

\newcommand{\del}[2]{\dfrac{d#1}{d#2}}

\newcommand{\lk}{\left}
\newcommand{\rk}{\right}

\newcommand{\HD}{H^{\dagger}}

\newcommand{\yc}{Y_{\chi}}
\newcommand{\ys}{Y_S}
\newcommand{\ycb}{\overline{Y}_{\chi}}
\newcommand{\ysb}{\overline{Y}_S}

\newcommand{\sigvcsm}{\sigma_v^{\chi\chi\rightarrow EE}}
\newcommand{\sigvcss}{\sigma_v^{\chi\chi\rightarrow SS}}
\newcommand{\sigvssm}{\sigma_v^{SS\rightarrow EE}}
\newcommand{\sigvscc}{\sigma_v^{SS\rightarrow\chi\chi}}

\newcommand{\sipC}{\sigma_{\chi, SI}}
\newcommand{\sipS}{\sigma_{S, SI}}

\newcommand{\mhez}{M^2_{H_1}}
\newcommand{\mhzz}{M^2_{H_2}}

\newcommand{\mhev}{M^4_{H_1}}
\newcommand{\mhzv}{M^4_{H_2}}

\newcommand{\ca}{\cos\alpha}
\newcommand{\sa}{\sin\alpha}
\newcommand{\caz}{\cos^2\alpha}
\newcommand{\saz}{\sin^2\alpha}

\newcommand{\gev}{\mathrm{GeV}}

\newcommand{\mh}{M_{H_2}}
\newcommand{\mhone}{M_{H_1}}
\newcommand{\mdm}{M_\chi}

\newcommand{\mpss}{\mu_{\phi SS}}

\title{\LARGE \bf A minimal model for two-component dark matter}
\author{Sonja Esch\footnote{sonja.esch@uni-muenster.de}, Michael Klasen\footnote{michael.klasen@uni-muenster.de}  ~and 
Carlos E. Yaguna\footnote{carlos.yaguna@uni-muenster.de} \\ 
\it \small Institut f\"ur Theoretische Physik, Universit\"at M\"unster,\\
\it \small Wilhelm-Klemm-Stra\ss e 9, D-48149 M\"unster, Germany}
\date{}  

\begin{document}
\maketitle
\vspace*{-7cm}
\begin{flushright}
\texttt{MS-TP-14-22}
\end{flushright}
\vspace*{6cm}
\begin{abstract}
We propose and study a new minimal model for two-component dark matter. The model contains only three  additional fields, one fermion  and two scalars,  all singlets under the Standard Model gauge group.  Two of these fields, one fermion and one scalar, are odd under a $Z_2$ symmetry that renders them simultaneously stable. Thus, both particles contribute to the observed dark matter density.  This model resembles the union of the singlet scalar and the singlet fermionic models but it contains some new features of its own. We analyze in some detail its dark matter phenomenology. Regarding the  relic density, the main novelty is the possible annihilation of one dark matter particle into the other, which can affect the predicted relic density in a significant way.  Regarding dark matter detection, we identify a new contribution that can lead either to an enhancement or to a suppression of the  spin-independent cross section for the scalar dark matter particle.  Finally, we define a set of five benchmarks models compatible with all present bounds and examine their direct detection prospects at planned experiments. A generic  feature of this model is that both particles give rise to observable signals in 1-ton direct detection experiments. In fact, such experiments will  be able to  probe even a subdominant dark matter component at the percent level.
\end{abstract}

\section{Introduction}
Current observations \cite{Ade:2013lta,Hinshaw:2012aka} indicate that most of the matter in the Universe consists of non-baryonic dark matter, but they do not tell us what this dark matter consists of. Since the Standard Model (SM), which has been extremely successful in describing all current collider data, does not contain any dark matter candidates,  the existence of dark matter provides strong evidence for  physics beyond the SM. If that new physics lies at the TeV scale (the scale that is currently being probed by the LHC), the observed dark matter density can be naturally obtained via the freeze-out mechanism in the early Universe --the so-called WIMP (Weakly Interacting Massive Particle) miracle.  Within this WIMP framework, several dark matter models have been studied, from those inspired by supersymmetry \cite{Jungman:1995df,Bertone:2004pz} or extra-dimensions \cite{Hooper:2007qk,Bertone:2004pz} to simpler  models that extend the  SM in a minimal way.

The idea behind minimal models of dark matter is to consider the simplest extensions of the SM that can account for the dark matter. In these models, the SM particle content is extended by a small number of fields, and a new discrete symmetry is usually introduced to guarantee the stability of the dark matter particle. Several variations can be obtained depending on the number and type of new fields (e.g. a scalar or a fermion, a singlet or a doublet under $SU(2)$, etc.) and on the discrete symmetry imposed ($Z_2$, $Z_3$, etc).  They include models such as the singlet scalar \cite{McDonald:1993ex, Burgess:2000yq}, the inert doublet \cite{LopezHonorez:2006gr,Barbieri:2006dq}, the singlet fermion \cite{LopezHonorez:2012kv}, higher scalar multiplets \cite{Hambye:2009pw}, minimal dark matter \cite{Cirelli:2005uq}, and $Z_N$ models \cite{Belanger:2012vp,Belanger:2012zr}, to name a few.  The main advantage of these models is that because they introduce only a small number of free parameters, they tend to be quite predictive. 

Even though it is often assumed that the dark matter density is entirely explained by a single particle, this is not necessarily the case. Two or even more particles could contribute to the observed dark matter density, a situation referred to as multi-component dark matter. This possibility has already  been considered  in a number of published works --see e.g.  \cite{Duda:2001ae,Duda:2002hf,Profumo:2009tb,Feldman:2010wy,Baer:2011hx,Aoki:2012ub,Bhattacharya:2013hva,Bian:2013wna,Kajiyama:2013rla}. In this paper we propose  a new model for two-component dark matter and we analyze its phenomenological implications. The most salient feature of this model is its simplicity. Besides the two dark matter particles (one scalar and one fermion, both SM singlets), the model contains only one additional field, a singlet scalar field that slightly mixes with the SM Higgs boson,  and  a single $Z_2$ symmetry is used to stabilize  both dark matter particles.  The model can be seen as the union of the singlet fermionic model \cite{LopezHonorez:2012kv,Kim:2008pp,Baek:2011aa,Baek:2012uj,Fairbairn:2013uta,Esch:2013rta} and the singlet scalar model \cite{McDonald:1993ex, Burgess:2000yq,Yaguna:2008hd,Goudelis:2009zz,Mambrini:2011ik}, but it has some new elements of its own. There are new processes affecting the relic density, including the annihilation of one type of dark matter into the other, which we study in some detail. There are also new contributions to the spin-independent direct detection cross section that can increase or decrease the predicted detection rate.  We examine the detection prospects of this model and show that both particles usually produce observable signals in planned direct detection experiments. In fact, those experiments can probe even a subdominant  dark matter component at the per cent level.

The rest of the paper is organized as follows. In the next section we introduce the model, discuss its main features, and determine its relevant parameter space.  In section \ref{sec:rd} we study how the relic density for both dark matter components depends on the parameters of the model. Special attention is paid to the role of dark matter conversion. Section \ref{sec:dd} deals with the direct detection cross sections. We provide the analytical results and numerically study a new contribution present in this model. In section \ref{sec:bench} we define a set of benchmark models that are compatible with all current bounds--including the relic density--and analyze their direct  detection prospects. Finally, we present our conclusions in section \ref{sec:conc}.  

\section{The model}
\label{sec:mod}
The model we propose is an extension of the Standard Model (SM) by three additional fields, two scalars and one fermion, all singlets under the gauge symmetry. Two of these new fields, one fermion ($\chi$) and one scalar ($S$), are assumed to be odd under a $Z_2$ symmetry that guarantees the stability of the lightest odd particle. All the  SM fields as well as the other scalar ($\phi$) are instead even under the $Z_2$.  Remarkably, in this setup the heavier odd particle turns out to be also  stable as there are no allowed interaction terms in the Lagrangian including both odd fields. In other words, the model has  an accidental symmetry that stabilizes  the heavier odd particle. Consequently, the model contains two dark matter particles.

The mass and interaction terms involving the dark matter fermion ($\chi$) are given by 
\begin{equation}
 \mathcal{L} = - \frac{1}{2}( M_{\chi} \overline{\chi} \chi + g_s \phi \overline{\chi} \chi + g_p \phi  \overline{\chi} \gamma_5 \chi )
\end{equation}
where $\phi$ is new the scalar field even under the $Z_2$ and $g_s$, $g_p$ are respectively the scalar and pseudoscalar couplings of $\chi$. Notice that, as anticipated, it is not possible to write interaction terms involving both $\chi$ and $S$ that are invariant under the gauge and the $Z_2$ symmetry.  

The scalar potential of this model can be written as 
\begin{align}
 V(\phi,H,S) &=  \mu_h^2(\HD H) + \lambda_H (\HD H)^2 - \frac{\mu_{\phi}^2}{2}\phi^2 + \frac{\lambda_{\phi}}{4}\phi^4  + \mu_1^3\phi \nonumber\\
 &+ \frac{\mu_3}{3}\phi^3 + \frac{\lambda_4}{2}\phi^2 (\HD H) 
                + \mu\,\phi(\HD H) - \frac 12\mu_S^2S^2+\frac{\lambda_S}{4}S^4 \nonumber\\
&+ \frac{\lambda}{4}(\HD H)S^2 + \frac{\mpss}{2}\phi S^2+\frac{\lambda_{\phi\phi SS}}{2}\phi^2S^2,
\label{eq:V}
\end{align}
where $S$ is the scalar dark matter particle and $H$ is the SM Higgs doublet that breaks the electroweak symmetry after acquiring a vacuum expectation value, $\langle H\rangle=\frac {1}{\sqrt 2} \left(\begin{smallmatrix} 0\\ v\end{smallmatrix}\right)$. Concerning $\phi$, it is always possible to choose a basis (by shifting the field) in such a way that $\langle\phi\rangle=0$, and so $\mu_1^3=-\mu v^2/2$. In the following, we will always work in that basis.

Due to the $\mu$ term in (\ref{eq:V}),   $h$ (the SM Higgs) and $\phi$ mix with each other giving rise to two scalar mass eigenstates, $H_1$ and $H_2$, defined as
\begin{equation}
 H_1= h\,\cos\alpha+\phi\,\sin\alpha,\quad H_2= \phi\,\cos\alpha  -h\,\sin\alpha,
\end{equation}
where $\alpha$ is the mixing angle.  We assume this mixing angle to be small so that we can identify $H_1$ with the  SM-like Higgs  observed at the LHC \cite{Aad:2012tfa,Chatrchyan:2012ufa} with a mass of about $125~\gev$. The other scalar, $H_2$, we take to be heavier, $\mh>\mhone=125~\gev$.

Even though the model introduces  $13$ new free parameters, not all of them are important to our discussion. The quartic couplings, $\lambda_\phi$ and $\lambda_S$, for instance, are irrelevant to  the dark matter  phenomenology.  And we can  take  $\mu_3=0$ and $\lambda_4=0$ without missing any critical effects. The remaining free parameters can be chosen to be
\begin{equation}
 \mdm, M_S,\mh,g_s,g_p,\sin \alpha, \lambda,\mpss, \lambda_{\phi\phi SS}.
\end{equation}
Most of these parameters can be associated with either the scalar or the fermionic dark matter sectors. $\mdm$, $g_s$,  and $g_p$ affect only the fermionic sector whereas $M_S$, $\lambda$, and $\lambda_{\phi\phi SS}$ concern only the scalar sector. Both sectors are influenced by $\mh$, $\sa$, and $\mpss$. In the next sections, we will study how the dark matter phenomenology of this model depends on these parameters.

As stated before, this model can be seen as  the union of the singlet fermionic model and the singlet scalar model, both of which have been previously studied --e.g. in \cite{LopezHonorez:2012kv,Kim:2008pp,Baek:2011aa,Baek:2012uj,Fairbairn:2013uta,Esch:2013rta} and \cite{McDonald:1993ex, Burgess:2000yq,Yaguna:2008hd,Goudelis:2009zz,Mambrini:2011ik}. It reduces to the singlet fermionic model in the absence of $S$ and to the singlet scalar model in the absence of $\chi$ and $\phi$. It contains, however, new terms and novel features not present in any of those two models. It is precisely these new features which are the main focus of this paper. 

\section{The relic density}
\label{sec:rd}

\begin{figure}[t!]
\begin{center}
\begin{tabular}{cc}
\includegraphics[width=0.4\textwidth]{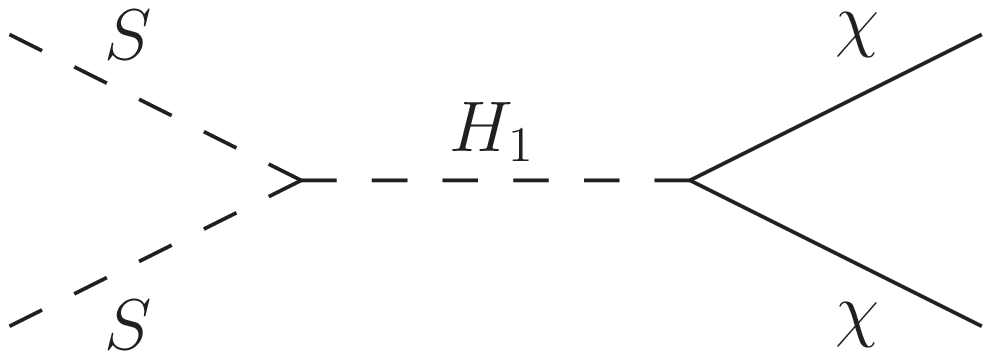} &
\includegraphics[width=0.4\textwidth]{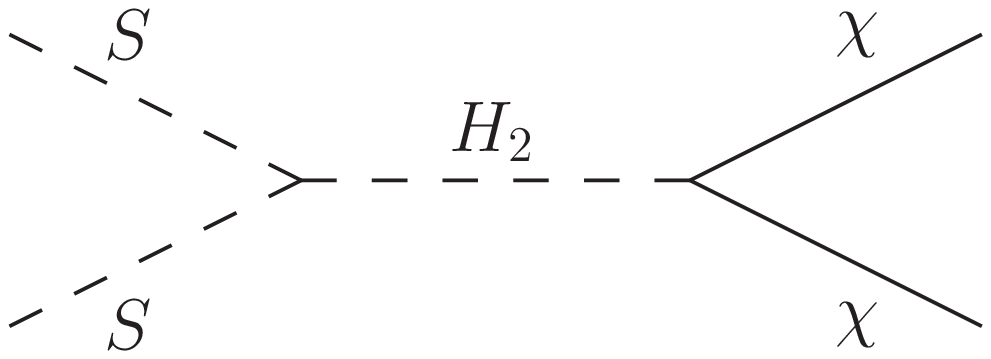}
\end{tabular}
\caption{\small\it The Feynman diagrams that contribute to the process of dark matter conversion, $\chi\chi\leftrightarrow SS$, in this model.}
\label{fig:conv}
\end{center}
\end{figure}
Since the model contains two dark matter particles, $\chi$ and $S$, we need to simultaneously follow their abundances in the early Universe. The Boltzmann equations are given by \cite{Belanger:2012vp}
\begin{align}
 \del{\yc}{x} &= -\sqrt{\frac{45}{\pi}}M_{Pl}g^{1/2}_{*}\frac{m}{x^2}\lk[\sigvcsm\lk(\yc^2-\ycb^2\rk)+\sigvcss\lk(\yc^2-\ycb^2\dfrac{\ys^2}{\ysb^2}\rk)\rk], \\
 \del{\ys}{x} &= -\sqrt{\frac{45}{\pi}}M_{Pl}g^{1/2}_{*}\frac{m}{x^2}\lk[\sigvssm\lk(\ys^2-\ysb^2\rk)+\sigvscc\lk(\ys^2-\ysb^2\dfrac{\yc^2}{\ycb^2}\rk)\rk],
\end{align}
where $E$ denotes any even particle (SM fermions or gauge bosons as well as $H_1$ and $H_2$),  $\sigma_v^{AA\rightarrow BB}$ is short for the thermally averaged annihilation cross section times velocity for the process $AA \to BB$, and $x = \frac{m}{T}$ with $T$ the temperature and $m = \frac{M_S + M_{\chi}}{2}$. In these equations $\overline Y$, $M_{Pl}$ and $g^{1/2}_*$ stand respectively for the equilibrium value of $Y$, the Planck mass and the degrees of freedom parameter. Besides the usual term accounting for dark matter annihilation into even particles (the first term), these equations describe also the conversion of one dark matter particle into the other, $\chi\chi\leftrightarrow SS$. Notice that, since $\sigvcss$ and $\sigvscc$ are determined by the same squared matrix element, they are not independent and are related to each other by
\begin{equation}
 \overline{Y}_\chi^2\sigvcss=\overline{Y}_S^2\sigvscc.
\end{equation}
These dark matter conversion processes are mediated by $H_1$ and $H_2$ as illustrated in figure \ref{fig:conv}. Since the coupling of $\chi$ to $H_1$ is suppressed by $\sin\alpha$, it usually is the $H_2$-mediated diagram that gives the dominant contribution --provided that $\mpss$ (which determines the $SSH_2$ vertex) is not too small. If $\chi$ and $S$ are close in mass the conversion can take place in both directions, $\chi\chi\to SS$ and $SS\to\chi\chi$, but if that is not the case only the conversion of the heavier particle into the lighter one is relevant.

To solve the above Boltzmann equations numerically, we have implemented the model into micrOMEGAs \cite{Belanger:2013oya}  (via LanHEP \cite{Semenov:2010qt}) and have used two different algorithms to integrate them. The first one is explained in \cite{Belanger:2012vp} and has already been incorporated into a new version of micrOMEGAs (not yet public) suited for models with two dark matter particles. The second one is of our own making and is based on the DarkSUSY \cite{Gondolo:2004sc} routines for the solution of the evolution equation in the case with only one dark matter particle. Even in this second case we relied on micrOMEGAs for the calculation of the relevant $\sigma_v$'s. We found that both procedures lead to the same values for the relic densities.

\begin{figure}[t!]
\begin{center}
\includegraphics[width=0.8\textwidth]{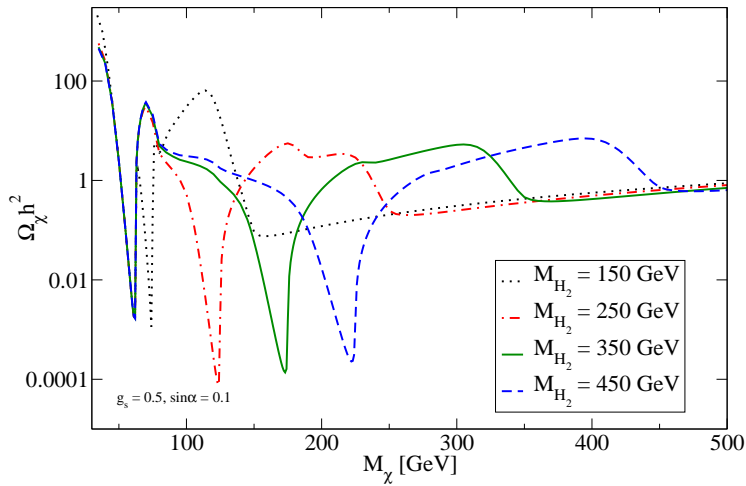} 
\caption{\small\it The $\chi$ relic density as a function of its mass for different values of $\mh$. In this figure we set $g_s=0.5$, $\sa=0.1$ and  $M_S=800~\gev$.  All other parameters were taken to be zero. }
\label{fig:rdmchidef}
\end{center}
\end{figure}

In this paper, we will only be concerned with freeze-out solutions to the relic density constraint. Freeze-in solutions also exist \cite{Hall:2009bx}, as both dark matter particles are SM singlets, but they require very small couplings and consequently do not give rise to any observable signals in dark matter experiments. Moreover, freeze-in was already studied both in the singlet scalar model \cite{Yaguna:2011qn} and in the singlet fermionic model \cite{Klasen:2013ypa} and we do not expect significant modifications to those results in our model. Thus, in the following we only examine regions in  the parameter space of this model where the dark matter particles have couplings large enough to reach thermal equilibrium in the early Universe so that their relic densities are the result of a freeze-out process.   

Since we have two stable particles, the dark matter constraint in this model reads
\begin{equation}
 \Omega_{DM}h^2=\Omega_Sh^2+\Omega_\chi h^2=0.1199\pm0.0027
\end{equation}
according to the data by PLANCK \cite{Ade:2013lta} and WMAP \cite{Hinshaw:2012aka}. A useful related quantity is   the fraction of the dark matter density that is due to $\chi$ and $S$, respectively denoted by $\xi_\chi$ and $\xi_S$. We have
\begin{equation}
 \xi_\chi=\frac{\Omega_{\chi}}{\Omega_{DM}},\quad \xi_S=\frac{\Omega_{S}}{\Omega_{DM}},\quad \mathrm{with}~~\xi_\chi+\xi_S=1.
\end{equation}
In this section, however, we will study the dependence of the relic density on the different parameters of the model without imposing this constraint. It will be taken into account in section \ref{sec:bench}, where the detection prospects will also be examined.

To begin with let us examine the fermion and scalar relic densities in the limit where dark matter conversion processes  ($\chi\chi\leftrightarrow SS$) are negligible. To that end we set $\mpss=0$ so that the $H_2$ mediated diagram in figure \ref{fig:conv} is suppressed. Figure \ref{fig:rdmchidef} shows the fermion relic density, $\Omega_\chi h^2$, as a function of $M_\chi$ for different values of $\mh$. For any given value of $\mh$ the relic density features a double dip at the $H_1$ and $H_2$ resonances (respectively at $M_\chi\sim 62.5~\gev, \mh/2$) and a marked decrease around $M_\chi\sim \mh$ due to the opening of the $\chi\chi\to H_2H_2$ annihilation channel. At high dark matter masses, the relic density is seen to increase with $M_\chi$ and to become independent of $\mh$. Regarding the final states from dark matter annihilation, they are dominated by gauge bosons ($W^+W^-$ and $Z^0Z^0$) for $M_\chi\lesssim \mh$ and by $H_2H_2$ for $M_\chi\gtrsim \mh$.  For the range of parameters illustrated in the figure, the relic density is seen to vary between $10^3$ at low dark matter masses and $10^{-4}$ at the $H_2$ resonance.

\begin{figure}[t!]
\begin{center}
\includegraphics[width=0.8\textwidth]{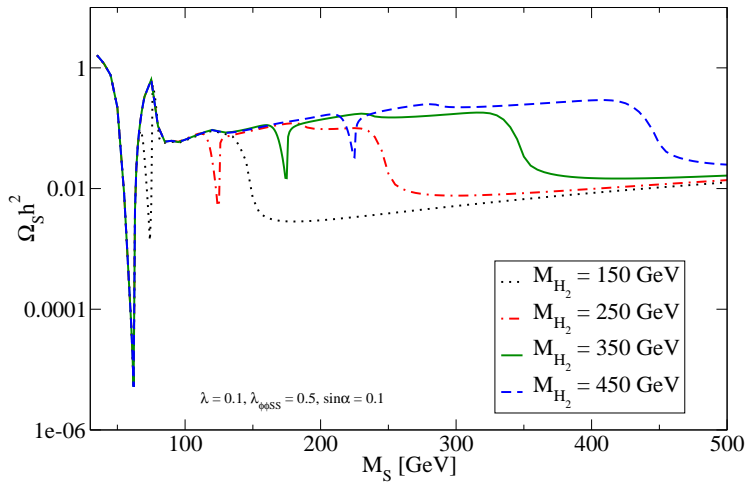} 
\caption{\small\it The $S$ relic density as a function of its mass for different values of $\mh$. In this figure we set $\lambda=0.1$, $\sa=0.1$, $\lambda_{\phi\phi SS}=0.5$ and $M_\chi=800~\gev$. All other parameters were taken to be zero.}
\label{fig:rdmsdef}
\end{center}
\end{figure}

The scalar relic density, $\Omega_Sh^2$, is shown in figure \ref{fig:rdmsdef} for the same values of $\mh$. In this figure, $\lambda$ and $\sa$ were set equal to $0.1$. A novelty in this model with respect to the singlet scalar is the existence of annihilations of the type $SS\to H_2H_2$. To illustrate their possible effect on the relic density we have set the parameter $\lambda_{\phi\phi SS}$ to $0.5$. In the figure we see that the relic density is very suppressed at the Higgs resonance ($M_S\sim 62.5~\gev$) but not so much at the $H_2$ resonance, in agreement with the fact that the $SSH_2$ coupling is suppressed by $\sa$. Notice also the significant decrease in the relic density at $M_S\sim \mh$, indicating that $SS\to H_2H_2$ becomes the dominant annihilation process in that region (at lower masses it is instead $SS\to W^+W^-$). It is important to stress though that, in contrast to the fermionic case, this final state is not always dominant for $M_S\gtrsim \mh$. Had we taken a small value of $\lambda_{\phi\phi SS}$, the only difference between the four lines would have been the position of the $H_2$ resonance.  Regarding the dependence on the mass, from the figure we see that the scalar relic density is largest at small masses and smallest at the $H_1$ resonance, and that it increases with $M_S$ at large masses.

\begin{figure}[t!]
\begin{center}
\includegraphics[width=0.8\textwidth]{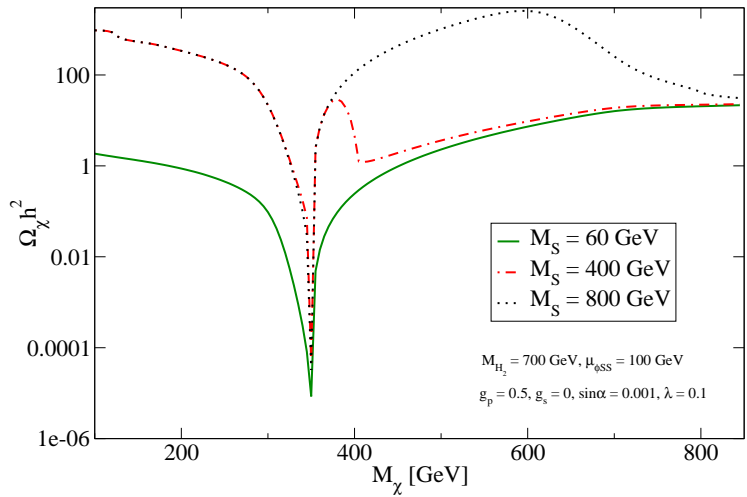} 
\caption{\small\it The effect of $\chi\chi\to SS$ on the $\chi$ relic density.  The lines correspond to different values of $M_S: 60~\gev, 400~\gev, 800~\gev$. The other relevant parameters were taken as $\mh=700~\gev$, $\mu_{\phi SS}=100~\gev$, $g_p=0.5$, $\sa=10^{-3}$.}
\label{fig:rdms}
\end{center}
\end{figure}

Let us next examine how  dark matter conversion can affect the predicted relic density. As discussed before, these processes are determined by the parameter $\mpss$ and are typically relevant in only one direction: the heavier dark matter particle annihilating into the lighter dark matter particle. Figure \ref{fig:rdms} shows the fermion relic density as a function of $M_\chi$ for different values of $M_S$: $60~\gev$, $400~\gev$ and $800~\gev$. For this figure we took $\mh=700~\gev$, $\mpss=100~\gev$, $\sa=10^{-3}$, $g_p=0.5$. The effect of the $H_2$ resonance ($M_\chi\sim 350~\gev$) is clearly visible in all three lines. When $M_S=60~\gev$, lower (solid) line, the annihilation channel $\chi\chi\to SS$ is kinematically open over the entire range of $M_\chi$, consequently the relic density is always smaller or equal than for the other two values of $M_S$. When $M_S=400~\gev$, the $\chi$ relic density is  larger over the range $M_\chi\lesssim M_S$, sharply decreases at $M_\chi\sim M_S$ and at higher dark matter masses it tends to the same value as for $M_S=60~\gev$. When $M_S=800~\gev$ the $\chi$ relic density coincides with that for $M_S=400~\gev$ for $M_\chi\lesssim 400~\gev$ or so. From that point on, it continues to increase for a while and it then decreases due to the opening of the $H_2H_2$ final state, joining the other two lines at high dark matter masses. Notice, in particular, that the effect of dark matter conversion gets significantly reduced for $M_\chi\gtrsim \mh$. From the figure we can see that, for the set of parameters chosen, the conversion of $\chi$ into $S$ ($\chi\chi\to SS$) may decrease $\Omega_\chi h^2$ by more than two orders of magnitude.

\begin{figure}[t!]
\begin{center}
\includegraphics[width=0.8\textwidth]{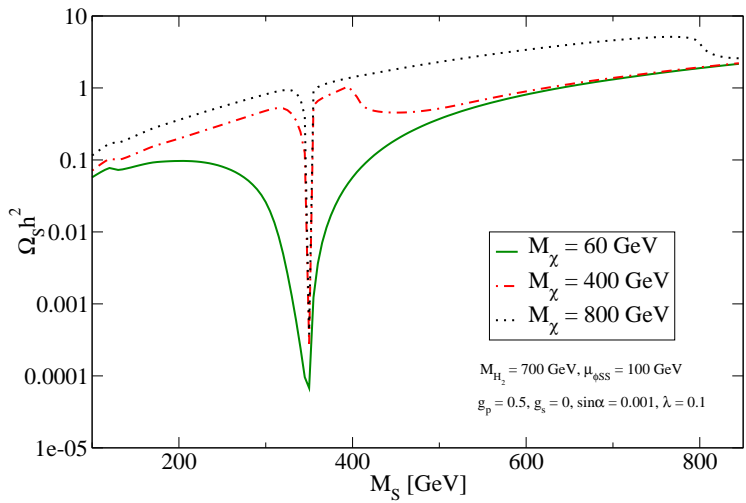} 
\caption{\small\it The effect of $SS\to\chi\chi$ on the $S$ relic density.  The lines correspond to different values of $M_\chi: 60~\gev, 400~\gev, 800~\gev$. 
The other relevant parameters were taken as $\mh=700~\gev$, $\mu_{\phi SS}=100~\gev$, $g_p=0.5$, $\sa=10^{-3}$.
}
\label{fig:rdmchi}
\end{center}
\end{figure}

If we now exchange the roles of $\chi$ and $S$ we obtain figure \ref{fig:rdmchi}, which displays the scalar relic density as a function of $M_S$ for different values of $M_\chi$.  All other parameters take the same values as in figure \ref{fig:rdms}. The relic density is smallest for $M_\chi=60~\gev$ (solid line), when $SS\to \chi\chi$ can take place over most of the $M_S$ range. For $M_\chi=400~\gev$ (dash-dotted line), we observe a sudden decrease in the relic density for $M_S\sim 400~\gev$ due to the opening of the $SS\to \chi\chi$ channel. Another effect caused by the processes $SS\leftrightarrow \chi\chi$ is the small difference observed at low $M_S$ between  the relic densities for $M_\chi=400~\gev$ and $M_\chi=800~\gev$ (dotted line). Naively one would not expect a heavy particle to affect the relic density of a lighter one, but, as illustrated in the figure, this seems to be the case. What happens in this example is that when $M_\chi=800~\gev$, $\chi$ freezes-out before $S$ and with a larger abundance.  While decoupled, $\chi$ has some residual annihilations into $S$, increasing its relic density. If instead $M_\chi=400~\gev$, $\chi$ freezes-out after $S$ and with a smaller abundance, and, therefore, the effect of residual annihilations is negligible. From the figure we see that the effect of $SS\to \chi\chi$ is significant over the entire mass range and it becomes more pronounced close to the $H_2$ resonance.

\begin{figure}[t!]
\begin{center}
\includegraphics[width=0.8\textwidth]{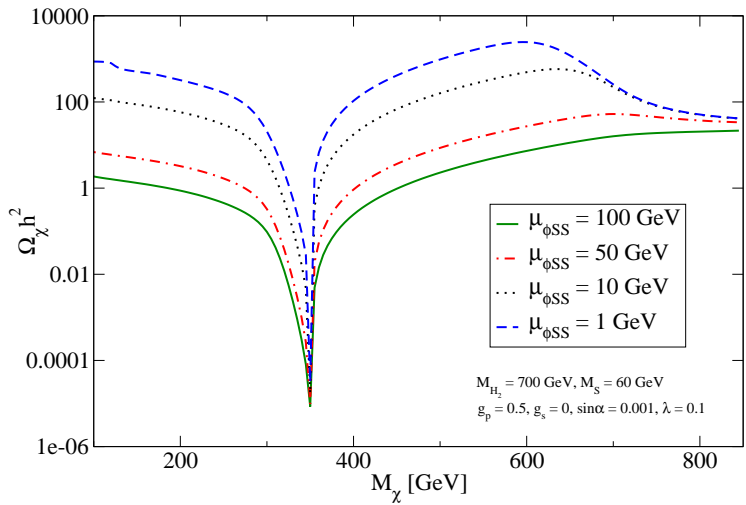} 
\caption{\small\it The dependence of the $\chi$ relic density on $\mu_{\phi SS}$. The lines correspond to $\mu_{\phi SS}= 100~\gev, 50~\gev, 10~\gev, 1~\gev$. The other relevant parameters were taken as $\mh=700~\gev$, $M_S=60~\gev$, $g_p=0.5$, $\sa=10^{-3}$.}
\label{fig:rdmu}
\end{center}
\end{figure}

The dependece of the $\chi$ relic density on $\mpss$ is illustrated in figure \ref{fig:rdmu}. In it we display $\Omega_{\chi}h^2$ as  a function of $M_\chi$ for $\mpss=100,\,50, 10\,,1\,\gev$. $M_S$ was set to $60~\gev$ and the other parameters were taken as in figure \ref{fig:rdms}. As expected, the larger $\mpss$ the larger the $\chi\chi\to SS$ annihilation rate and therefore the smaller the relic density. In the figure we observe that the value of $\Omega_\chi$ can differ by more than two orders of magnitude between $\mpss=100~\gev$ and $\mpss=1~\gev$. Notice that for $\mpss=100~\gev$ (solid line) the process $\chi\chi\to SS$ is so dominant that the behavior of the relic density does not even change at $M_\chi\sim M_{H_2}=700~\gev$, where the  annihilation channel $H_2H_2$ opens up. But if $\mpss=10~\gev$ (dotted line) or $\mpss=1~\gev$ (dashed line) the relic density does get reduced around $M_\chi\sim 700~\gev$, implying that the $H_2H_2$ annihilation channel becomes relevant. As a result, for $M_\chi\gtrsim 700~\gev$ the variation of the relic density for different values of $\mpss$ is not so large.

To summarize, the main novelties of  this model with respect to the singlet scalar and the singlet fermionic models regarding the dark matter relic densities are the presence of a new resonance (at $M_S\sim M_{H_2}/2$) and of a new final state ($H_2H_2$) that can affect the relic density of the scalar, and the  conversion of one dark matter particle into the other, which may reduce the relic density of the heavier and increase that of the lighter, modifying in a significant way the viable regions. Next, we will briefly review how  dark matter detection is modified in this model.

\section{Direct detection}
\label{sec:dd}
Direct detection is probably the most promising way of detecting dark matter. On the theoretical side, it is subject to fewer astrophysical uncertainties than indirect detection and so the  predictions tend to be more reliable. On the experimental side, direct detection experiments such as XENON100 \cite{Aprile:2012nq} and LUX \cite{Akerib:2013tjd} have already made outstanding progress during the last few years and have set strong bounds on the dark matter spin-independent direct detection cross section. In addition, planned 1-ton experiments such as XENON1T, which should start taking data next year, are expected to improve the current sensitivity by about two orders of magnitude. Thus, they have great chances of discovering the dark matter particle. 

In our model, both spin-independent direct detection cross sections tend to be rather large, as evidence by the fact that in the singlet scalar model and in the singlet fermionic model  significant regions of the parameter space are already excluded by present bounds \cite{Esch:2013rta,Mambrini:2011ik}. Most of these bounds change little when the two models are combined, as we explain in the following. Since $\chi$ and $S$ account only for a fraction $\xi_\chi$ and $\xi_S$ of the dark matter density, the quantities that are actually constrained by direct detection experiments are $\xi_\chi\sipC$ and $\xi_S\sipS$, where $\sipC$ and $\sipS$ are the usual spin-independent cross sections for the fermion and the scalar dark matter particles. This does not mean, however, that the detection rates are suppressed by $\xi$. As noted sometime ago \cite{Duda:2001ae}, the direct detection rate does not strongly depend on $\xi$. The reason is simple to understand: a smaller fraction of the relic density would generally imply  a larger dark matter coupling (say  a larger $\lambda$ or $g_s$) that would in turn translate into a larger spin-independent cross section, exactly canceling out the $\xi$ suppression in the direct detection rate. On the other hand, the indirect detection rate, $\sigma v$, has to be multiplied by $\xi^2$, which does leave a $\xi$ suppression after taking into account the larger coupling required to reduce $\Omega$. Thus, models with multi-component dark matter generally have better detection prospects in direct detection experiments. 

\begin{figure}[t!]
\begin{center}
\begin{tabular}{cc}
\includegraphics[width=0.4\textwidth]{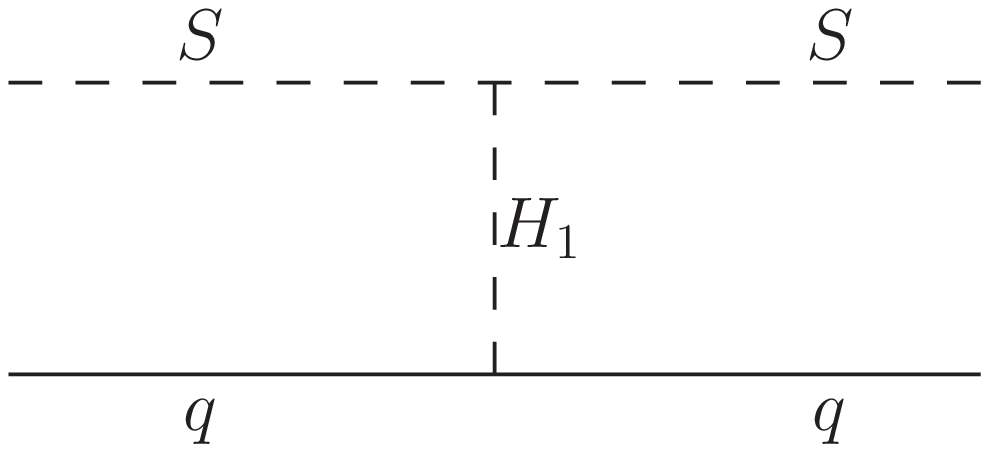} &
\includegraphics[width=0.4\textwidth]{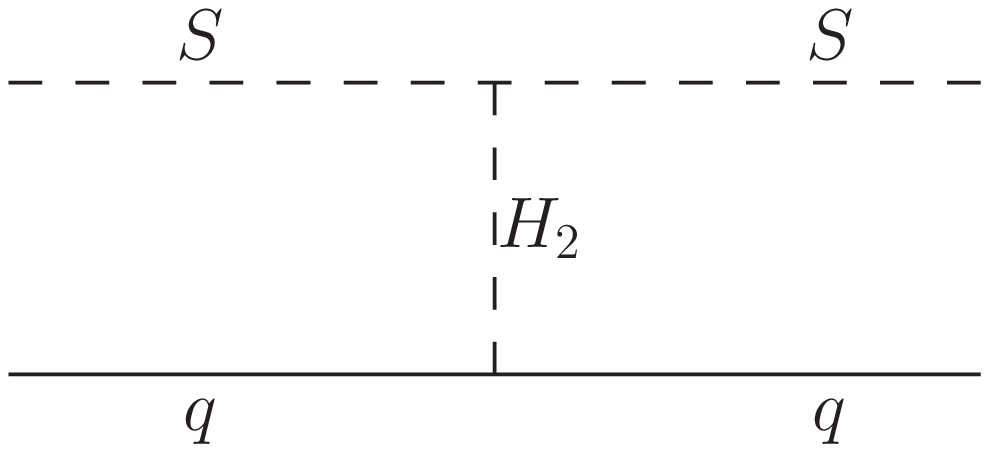}
\end{tabular}
\caption{\small\it The Feynman diagrams that contribute to the elastic scattering off quarks for the scalar dark matter particle  in this model.}
\label{fig:dd}
\end{center}
\end{figure}

The spin-independent direct detection cross section of $\chi$ is determined by  $H_1$ and $H_2$ mediated diagrams  and  is the same as in the singlet fermionic model,
\begin{equation}
 \sipC=\frac{g_s^2\sin^22\alpha}{4\pi}m_r^2\left(\frac{1}{\mhez}-\frac{1}{\mhzz}\right)^2 g_{Hp}^2,
\label{eq:sipc}
\end{equation}
where $m_r$ is the reduced mass and 
\begin{equation}
 g_{Hp}=\frac{m_p}{v}\left[\sum_{q=u,d,s}f_q^p+\frac 29\left(1-\sum_{q=u,d,s}f_q^p\right)\right]\approx 1.4\times 10^{-3}.
\end{equation}
Thus, the only model parameters that enter into its evaluation are $\sa$, $g_s$, $M_\chi$ and $M_{H_2}$.

For the scalar, the spin-independent cross section has a new $H_2$-mediated contribution (see figure \ref{fig:dd}) not present in the singlet scalar model. The total cross section is given as 
\begin{align}
 \sipS &= \frac{m_r^2}{4\pi\mhev\mhzv M_S^2} \lk[\lambda\frac{v}{2} (\caz~\mhzz + \saz~\mhez)\rk.\nonumber\\
 &~~ + \lk.\mpss \ca \sa (\mhez - \mhzz)\rk]^2 g_{Hp}^2.
\label{eq:sips}
\end{align}
Hence, it is determined by  the parameters $M_S$, $\lambda$, $\sa$, $\mh$ and $\mpss$. The second term, in fact, is proportional to $\mpss$, the same parameter that controls dark matter conversion ($\chi\chi\leftrightarrow SS$), as we saw in the previous section. Interestingly, the two terms may cancel against each other, giving a suppressed cross section, when $\mpss$ takes the following value  
\begin{equation}
 \mpss^\mathrm{cancel} = \frac{\lambda v}{2}\lk( \frac{\ca}{\sa} + \frac{\mhez}{(\mhzz-\mhez)\ca\sa}\rk),
\end{equation}
which does not depend on the mass of the dark matter particle. Let us now study numerically this spin-independent cross section and, in particular, the cancellation effect. 

\begin{figure}[t!]
\begin{center}
\includegraphics[width=0.8\textwidth]{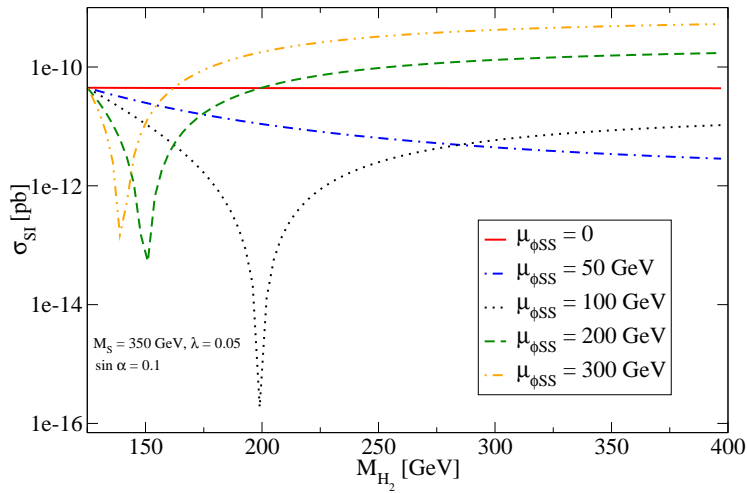} 
\caption{\small\it The scalar spin-independent cross section as a function of $\mh$ for different values of $\mpss$: $300~\gev$, $200~\gev$, $100~\gev$, $50~\gev$, $0$. In this figure, $M_S=350~\gev,\sa=0.1,\lambda=0.05$. }
\label{fig:sipscalar}
\end{center}
\end{figure}

Figure \ref{fig:sipscalar} shows $\sipS$ as a function of $\mh$ for different values of $\mpss$. The other parameters were chosen as $M_S=350~\gev$, $\lambda=0.05$ and $\sa=0.1$. When $\mpss=0$ (solid line), the dependence of $\sipS$ on $\mh$ disappears --see equation (\ref{eq:sips})-- and we simply obtain a constant. This constant is the reference value against which we are going to compare the behavior of $\sipS$ for $\mpss\neq 0$. If $\mpss=50~\gev$ (dash-dotted line), for instance, we see that we obtain a smaller value of $\sipS$ over the entire range of $\mh$. For $\mpss=100~\gev$ (dotted-line) we observe a strong suppression of $\sipS$ at $\mh\sim 200~\gev$ due to the above mentioned cancellations. For $\mpss=200,300~\gev$, this cancellation occurs instead for $\mh$ around $150~\gev$. In addition, notice that for those two values of $\mpss$, $\sipS$ is actually enhanced over most of the $\mh$ range. Thus, a non-zero value of $\mpss$ can lead either to an  increase or a decrease of the spin-independent cross section.

\begin{figure}[t!]
\begin{center}
\includegraphics[width=0.8\textwidth]{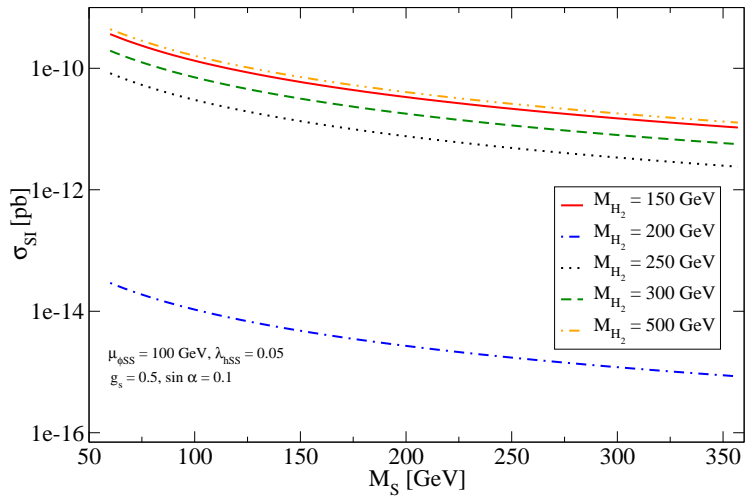} 
\caption{\small\it The scalar spin-independent cross section as a function of $M_S$ for different values of $\mh$.  In this figure, $\mpss=100~\gev,\sa=0.1,\lambda=0.05$. }
\label{fig:sipscalar2}
\end{center}
\end{figure}

In figure \ref{fig:sipscalar2} we display instead $\sipS$ as a function of $M_S$ for different values of $\mh$ and given values of $\mpss$ ($100~\gev$), $\sa$ (0.1), and $\lambda$ ($0.05$). The dependence with $\mh$ is clearly non-trivial. For $\mh=200~\gev$, $\sipS$ is highly suppressed  as this set of parameters satisfies the cancellation condition --see the dotted line in figure \ref{fig:sipscalar}. Since this condition is independent of $M_S$, the suppression holds for the entire mass range. For $\mh=150~\gev$ (solid line) and $\mh=500~\gev$ (dash-double dotted line), the value of $\sipS$ is practically the same and it is larger than that for $\mh=250~\gev$ (dotted line) and $\mh=300~\gev$ (dashed line).  We also notice that, as expected from equation \ref{eq:sips}, $\sipS$ decreases with $M_S$.

As we have seen, the spin-independent cross section of $\chi$ is identical to that found in the singlet fermionic model, whereas that of  $S$ receives a new contribution that can increase or decrease the detection rate. To compare the predicted spin-independent cross sections with the sensitivity of current and future experiments, we must first impose the dark matter constraint so that they are calculated only for models consistent with the observed dark matter density. That is what we do in the next section.

\section{Detection prospects}
\label{sec:bench}

\begin{figure}[t!]
\begin{center}
\includegraphics[width=0.8\textwidth]{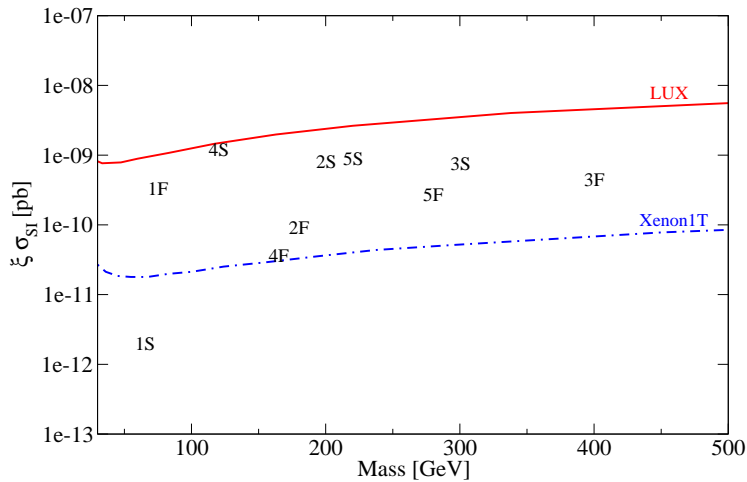} 
\caption{\small\it The location of the five benchmark models in the plane ($\xi\sigma$, $M$). The solid line shows the current bound from the LUX experiment \cite{Akerib:2013tjd}, while the dashed line shows the expected sensitivity of XENON1T \cite{Aprile:2012zx}.}
\label{fig:bench}
\end{center}
\end{figure}

To assess the dark matter detection prospects, we have selected a sample of five benchmarks models (or parameter space points) that are consistent with all phenomenological and cosmological bounds and that illustrate some of the interesting possibilities that can occur in this model.  For definiteness,  in this section we set $g_p=0$ (parity conserving case) and $\lambda_{\phi\phi SS}=0$. Our benchmarks are listed in table \ref{table}. For each model, we include the input parameters, the predicted relic densities, the spin-independent cross section, and, for completeness, the indirect detection rate $\sigma v$.  They are also displayed,  in the plane (Mass, $\xi\sigma_{SI}$), in figure \ref{fig:bench}, where we also compare their expected direct detection rates with the current bound from the LUX experiment \cite{Akerib:2013tjd}  and with the expected sensitivity of XENON1T \cite{Aprile:2012zx}. Let us now describe in detail each of these benchmark points.
\begin{table}
\begin{center}
\renewcommand{\arraystretch}{1.4}
\begin{tabular}{@{}l l  l l l l@{}}
\noalign{%
\setlength{\arrayrulewidth}{1.3pt}%
\let\noalign\empty
\hline
}%
Parameters / Model     & I & II & III & IV & V   \\
\noalign{%
\setlength{\arrayrulewidth}{1pt}%
\let\noalign\empty
\hline
}%
M$_{S}$~[GeV]                          &  65   & 200   & 300  & 120  & 220  \\
M$_{\chi}$~[GeV]                      &  75   & 180   & 400  & 165  & 280  \\
M$_{H_2}$~[GeV]                      &  250  & 150   & 200  & 360  & 250  \\
$\mpss$~[GeV]             & 400   & 0     & 0    & 0    & 0    \\
g$_s$                                     &  0.45 & 0.58  & 0.9  & 0.65  & 0.6 \\
$\sin\alpha$                            & 0.1   & 0.1   & 0.07  & 0.08  & 0.05  \\
$\lambda$                                 & 0.25   & 0.175 & 0.25 & 0.09 & 0.5  \\
\noalign{%
\setlength{\arrayrulewidth}{1pt}%
\let\noalign\empty
\hline
}%
$\Omega_S / \Omega_{DM}$ \hspace{0.1cm}[$\%$]            & 52 & 51 & 49 & 97 & 8  \\
$\Omega_{\chi} / \Omega_{DM}$ \hspace{0.1cm}[$\%$]       & 48 & 49 & 51 & 3  & 92 \\
\noalign{%
\setlength{\arrayrulewidth}{1pt}%
\let\noalign\empty
\hline
}%
$\sigma_{SI,S}$  \hspace{0.15cm}[pb]               & 2.9$\times$10$^{-12}$  & 1.7$\times$10$^{-9}$  & 1.5$\times$10$^{-9}$ & 1.20$\times$10$^{-9}$ & 1.1$\times$10$^{-8}$ \\
$\sigma_{SI,\chi}$ \hspace{0.15cm}[pb]             & 6.6$\times$10$^{-10}$  & 1.8$\times$10$^{-10}$ & 8.7$\times$10$^{-10}$ & 1.2$\times$10$^{-9}$ & 3.0$\times$10$^{-10}$ \\
\noalign{%
\setlength{\arrayrulewidth}{1pt}%
\let\noalign\empty
\hline
}%
$\sigma v_S$ \hspace{0.1cm}[10$^{-26}$cm$^3$/s]        & 7.2 & 4.7 & 4.5 & 2.2 & 30 \\
$\sigma v_{\chi}$ \hspace{0.1cm}[10$^{-26}$cm$^3$/s]  & 1.6$\times$10$^{-8}$ & 4.5$\times$10$^{-5}$ & 4.2$\times$10$^{-5}$ & 2.7 & 2.1$\times$10$^{-5}$ \\
\noalign{%
\setlength{\arrayrulewidth}{1.3pt}%
\let\noalign\empty
\hline
}%
\end{tabular} 
\label{fixedpoint}
\caption{The five benchmark models that we use to assess the dark matter detection prospects. Notice that $\sigma_{SI}$ and $\sigma v$ should be rescaled by the corresponding factors of $\xi$ and $\xi^2$.}
\label{table}
\end{center}
\end{table}

\paragraph{Model I}
In this  model both dark matter particles are very light, $M_S=65~\gev$, $M_\chi=75~\gev$, and give about the same contribution to the dark matter density.   Such low mass models are usually highly constrained--if not excluded altogether-- by direct detection bounds \cite{Esch:2013rta}. What allows this model to evade those constraints is the high value of $\mpss$ ($400~\gev$), which  permits the efficient annihilation of $\chi\chi$ into $SS$--so that $g_s$ does not have to be large to satisfy the dark matter constraint-- and at the same time suppresses $\sipS$ via the cancellation effect studied in the previous section. Indeed, had we set $\mpss=0$, the fermion relic density, $\Omega_\chi h^2$, would have increased to about $10$ and the scalar spin-independent cross section would have reached $10^{-8}$pb. The suppression of $\sipS$ is so effective that the scalar cross section lies well below the expected sensitivity of XENON1T. The fermion instead should be easily detected by future experiments.

\paragraph{Model II}
In this model the  dark matter particles have masses in the intermediate range ($200~\gev$ and $180~\gev$) and each  gives about a  $50\%$ contribution to the dark matter density. Since $\mh=150~\gev<M_\chi$, the fermion relic density is obtained mainly via annihilation into $H_2H_2$. Due to the closeness  between $\mh$ and $\mhone$, a slight suppression in the fermion spin-independent cross section is expected --see equation (\ref{eq:sipc}). From the figure we see that the scalar cross section is about one order of magnitude larger for the scalar than for the fermion and that  both dark matter particles could be detected in the next generation of experiments.

\paragraph{Model III}
In this model both dark matter particles are relatively heavy ($300~\gev$ and $400~\gev$) and each accounts for $50\%$ of the dark matter density. The annihilation of $\chi$ is dominated by the $H_2H_2$ final state and that of $S$ by the usual $W^+W^-$. Since $\mpss=0$, dark matter conversion processes play no role in this case.  The resulting spin-independent cross sections are between $10^{-9}$ and $10^{-10}$ pb, being slightly larger for the scalar, and are well within the expected sensitivity of XENON1T.

\paragraph{Model IV}
In this model the dark matter density is dominated by the scalar ($97\%$), which  has a spin-independent cross section close to the current LUX bound. The fermion has a small relic density--it accounts only for about $3\%$ of the dark matter-- due to the enhancement in its annihilation rate that occurs close to the $H_2$ resonance, $M_\chi\sim \mh/2$.  Even though $\chi$ gives a small contribution to the dark matter density, it may be observed in 1-ton  direct detection experiments. That is, such experiments can  probe not only the dominant dark matter component but also a subdominant one at the percent level.

\paragraph{Model V}  
In this case, both dark matter particles have intermediate masses ($220~\gev$ and $280~\gev$) with the fermion accounting for $92\%$ of the dark matter and the scalar for the remaining $8\%$. The suppression in the $S$ relic density is obtained simply by increasing the value of $\lambda$. In the figure we see that both dark matter particles have a spin-independent cross section well within the expected sensitivity of planned experiments. Again, even the subdominant dark matter component will be probed by such experiments.

Notice also from table \ref{table} that the indirect detection rate of the fermion, $\sigma v_{\chi}$, is always very small, even before taking into account the additional $\xi^2$ suppression. This is due to the fact that the annihilation rate is velocity dependent ($\sigma v\sim v^2$) and $v\sim 10^{-3}$ for dark matter particles in the Galactic halo.  $\sigma v_{S}$, on the other hand, is not highly suppressed but once multiplied by the corresponding factor $\xi_S^2$ it  always gives a value below the thermal one ($3\times 10^{-26}\mathrm{cm}^3/\mathrm{s}$). Hence, as anticipated, the indirect detection prospects in this model are not promising. 

\section{Conclusions}
\label{sec:conc}
We have proposed and analyzed a new minimal model for two-component dark matter. The model is an extension of the SM by three fields, one fermion and two scalars, that are singlets under the gauge group. Two of these fields, one fermion and one scalar, are odd under a $Z_2$ symmetry that automatically renders stable the lightest of the two. A nice feature of this model is that the heavier odd particle is also stable due to an accidental symmetry so that both particles contribute to the observed dark matter density.  The model can be seen as the union of the singlet fermionic model and the singlet scalar model, but it contains interesting new features of its own. The relic density, for example, is affected by the conversion of one dark matter particle into the other ($SS\leftrightarrow \chi\chi$), an effect we examined in some detail. There is also a new contribution to the scalar spin-independent cross section that could lead either to an enhancement or to a reduction of the predicted cross section. To assess the dark matter detection prospects in this model, we selected five benchmark points compatible with all phenomenological and cosmological constraints and computed their spin-independent cross sections and their indirect detection rates. We found that in most cases, both dark matter particles could be detected in planned direct detection experiments such as XENON1T. 

\section*{Acknowledgments}
We would like to thank Alexander Pukhov and Genevi\`eve B\'elanger for providing us with a not-yet-public version of micrOMEGAs suited for models with two dark matter particles. This work is partially supported by the ``Helmholtz Alliance for Astroparticle Physics HAP'' funded by the Initiative and Networking Fund of the Helmholtz Association.

\bibliographystyle{hunsrt}
\bibliography{darkmatter}

\end{document}